\documentclass[aip,pof,reprint]{revtex4-1}

\usepackage{amsmath, amssymb, amsfonts, amsthm}
\usepackage{bm}
\usepackage{geometry}
\usepackage{graphicx}
\usepackage{mathrsfs}

\usepackage{mathtools}
\usepackage{xcolor}
\usepackage[
  colorlinks=true,
  linkcolor=blue,
  citecolor=blue,
  urlcolor=cyan
]{hyperref}
\usepackage{cleveref}

\usepackage{tikz}
\usepackage{pgfplots}
\pgfplotsset{compat=1.18}
\usepackage{pgfplotstable}
\usetikzlibrary{arrows.meta, decorations, positioning, calc, fit, backgrounds, shapes.geometric}

\newcommand{\Vol}{\mathscr{V}}
\newcommand{\uvec}{\mathbf{u}}
\newcommand{\nvec}{\mathbf{n}}
\newcommand{\Bvec}{\mathbf{B}}
\newcommand{\Avec}{\mathbf{A}}
\newcommand{\Rvec}{\mathbf{R}}
\newcommand{\Jvec}{\mathbf{J}}
\newcommand{\Zvec}{\mathbf{Z}}

\begin{document}

\title{A Variational Formulation of the MHD Induction Equation}
\author{A. Farooq}
\email{afarooq@unb.ca}
%\affiliation{University of New Brunswick, Fredericton}

\begin{abstract}
We present a variational formulation of the induction equation in magnetohydrodynamics (MHD) based on Gauss's principle of least constraint. The central result is the Euler--Lagrange equation $\Zvec_B = -\nabla h_m$, where $h_m = \Avec\cdot\Bvec$ is the magnetic helicity density. This reveals that the magnetic helicity gradient $\nabla h_m$ acts as the constraint force maintaining the solenoidality of the magnetic field, exactly as the pressure gradient $\nabla p$ maintains incompressibility in Taha et al.'s pressure-gradient minimization principle. The magnetic helicity density—which, though gauge-dependent locally, yields a gauge-invariant variational principle—with magnetic helicity naturally emerging as the Lagrange multiplier enforcing $\nabla\cdot\Bvec=0$, and at the solution the field minimizes the norm of the magnetic helicity gradient $\|\nabla h_m\|^2$.  This establishes a structural analogy: magnetic helicity is to the magnetic field as pressure is to velocity. The variational principle connects to Woltjer's theorem, Taylor relaxation, and the Hamiltonian structure of MHD.
\end{abstract}

\maketitle

\section{Introduction}

Variational principles have played a central role in the development of magnetohydrodynamics (MHD), from the early work of Woltjer \cite{Woltjer1958} on force-free magnetic fields to the modern geometric formulations of Morrison and Greene \cite{MorrisonGreene1980}. Woltjer demonstrated that the magnetic energy is minimized subject to the constraint of constant global magnetic helicity, yielding the force-free condition $\nabla\times\Bvec = \mu\Bvec$. Taylor \cite{Taylor1974} extended this to plasmas, arguing that helicity decays much slower than energy during relaxation, leading to the Taylor state.

Recently, Taha et al. \cite{Taha2023} demonstrated that the incompressible Navier--Stokes equations can be derived from a variational principle in which the pressure $p$ acts as a Lagrange multiplier enforcing the incompressibility constraint $\nabla\cdot\uvec=0$, with the flow minimizing the pressure gradient. This pressure-gradient minimization principle (PGMP) has provided a new variational foundation for fluid dynamics.

In a companion paper \cite{Farooq2026}, we extended this framework to the vorticity equation, showing that the helicity density $h = \uvec\cdot\boldsymbol{\omega}$ emerges as the Lagrange multiplier enforcing the solenoidal constraint $\nabla\cdot\boldsymbol{\omega}=0$, with the flow minimizing the helicity gradient. In this work, we develop a variational principle for the induction equation of MHD with again the magnetic helicity density $h_m = \Avec\cdot\Bvec$ naturally emerging as the Lagrange multiplier enforcing the solenoidality of the magnetic field $\nabla\cdot\Bvec=0$. 

The paper is organized as follows. In Section 2,  we derive the variational principle and establish the central result $\Zvec_B = -\nabla h_m$. In Section 3, we discuss the physical interpretation, including the connection to Woltjer's theorem, Taylor relaxation, and magnetic reconnection. In Section 4, we compare with Taha's PMPG and the vorticity formulation. In Section 5, we offer an conclusion and discuss future directions. 

\section{A Variational Principle for the Induction Equation}

The induction equation governing the evolution of the magnetic field $\Bvec$ in resistive MHD is:
\begin{equation}
\Bvec_t = \nabla \times (\uvec\times\Bvec) - \nabla \times (\eta \nabla\times\Bvec)
\end{equation}

where $\uvec$ is the plasma velocity and $\eta$ is the magnetic diffusivity (resistivity). In the ideal limit ($\eta=0$), the induction equation reduces to the frozen-flux condition.

The solenoidal constraint on the magnetic field is given as $\nabla\cdot\Bvec=0$. The magnetic vector potential $\Avec$ is defined by $\Bvec = \nabla\times\Avec$. The magnetic helicity density is:
\begin{equation}
h_m := \Avec\cdot\Bvec
\end{equation}
The global magnetic helicity is:
\begin{equation}
H_m := \int_\Omega h_m \, d\Vol = \int_\Omega \Avec\cdot\Bvec \, d\Vol
\end{equation}
In ideal MHD ($\eta=0$), $H_m$ is a topological invariant (Woltjer\cite{Woltjer1958}) that measures the knottedness and linkage of magnetic flux tubes.

We now propose a variational principle for the induction equation. Following Gauss's principle of least constraint (see \cite{Lanczos2012}), we hypothesize that the field minimizes the squared norm of some constraint force, subject to the solenoidal constraint $\nabla\cdot\Bvec=0$.

A subtle but important point must be clarified regarding this constraint. While $\nabla\cdot\Bvec = \nabla\cdot(\nabla\times\Avec) \equiv 0$ is a strict kinematic identity for any vector field derived from a vector potential, in the variational calculus we treat $\Bvec_t$ as an independent field to be varied. The space of admissible variations $\delta\Bvec_t$ may contain trial fields that momentarily violate solenoidality. The constraint thus acts to restrict these variations, ensuring the dynamics remain on the divergence-free manifold—precisely analogous to the role of the pressure in maintaining $\nabla\cdot\uvec=0$ in the primitive formulation.

\subsection{Euler--Lagrange Equation}

Consider the unconstrained vector field:
\begin{equation}
\begin{split}
\Zvec_B := \Bvec_t 
&- \nabla \times (\uvec\times\Bvec) + \nabla \times (\eta \nabla\times\Bvec) \\
&- (\Avec\cdot\nabla)\Bvec - (\Bvec\cdot\nabla)\Avec - \Avec \times (\nabla \times \Bvec)
\end{split}
\label{eq:ZB_def}
\end{equation}
which represents the magnetic dynamics before the solenoidal constraint is imposed. This combination of terms may appear arbitrary at first glance. The appearance of the vector potential $\Avec$ is particularly curious, as it suggests a gauge-dependent quantity. Their meaning will become clear shortly.

We define the constraint Gaussian functional (see \cite{Lanczos2012}) as:
\[
\mathscr{G}_B = \frac{1}{2}\int_\Omega \|\Zvec_B\|^2 \, d\Vol
\]

We introduce the Lagrangian:
\begin{equation}
\mathscr{L}_B[\Bvec, \Bvec_t, \lambda] = \frac{1}{2} \int_\Omega \|\Zvec_B\|^2 \, d\Vol - \int_\Omega \lambda \, \nabla\cdot\Bvec_t \, d\Vol
\label{eqn:lagrangian_B}
\end{equation}
where $\lambda$ is a Lagrange multiplier enforcing $\nabla\cdot\Bvec_t=0$.

Taking the variation with respect to the independent field $\Bvec_t$ gives:
\begin{equation}
\delta_{\Bvec_t}\mathscr{L}_B
= \int_\Omega \Zvec_B \cdot \delta\Bvec_t \, d\Vol - \int_\Omega \lambda \, \nabla\cdot(\delta\Bvec_t) \, d\Vol
\end{equation}
since $\Zvec_B$ depends linearly on $\Bvec_t$. Applying the product rule $\nabla\cdot(\lambda \,\delta\Bvec_t) = \nabla\lambda \cdot \delta\Bvec_t + \lambda \nabla\cdot(\delta\Bvec_t)$ allows us to rewrite the second term in the variation as:
\begin{equation}
\int_\Omega \lambda \, \nabla\cdot(\delta\Bvec_t) \, d\Vol
=  \int_\Omega \nabla\cdot(\lambda \,\delta\Bvec_t) \, d\Vol
- \int_\Omega \nabla\lambda \cdot \delta\Bvec_t \, d\Vol
\end{equation}

The first term on the right is a total divergence. Applying the divergence theorem converts it to a surface integral over the boundary $\partial\Omega$:
\begin{equation}
\int_\Omega \nabla\cdot(\lambda \,\delta\Bvec_t) \, d\Vol
= \oint_{\partial\Omega} \lambda \, \delta\Bvec_t \cdot \nvec \, dS
\end{equation}
where $\nvec$ is the outward unit normal. Assembling the full variation gives:
\begin{equation}
\delta_{\Bvec_t}\mathscr{L}_B
= \int_\Omega (\Zvec_B + \nabla \lambda) \cdot \delta\Bvec_t \, d\Vol
- \oint_{\partial\Omega} \lambda \, \delta\Bvec_t \cdot \nvec \, dS
\end{equation}

The surface integral vanishes because, by the standard assumption of variational calculus, the admissible variations $\delta\Bvec_t$ vanish on the boundary $\partial\Omega$. This is a standard requirement and defines the domain of validity of the principle. With the surface term eliminated, stationarity of the action requires the volume integral to vanish for all admissible variations $\delta\Bvec_t$. This yields the Euler--Lagrange equation pointwise:
\begin{equation}
\Zvec_B = -\nabla \lambda
\label{eq:EL_lambda_B}
\end{equation}

\subsection{Identification of the Lagrange Multiplier}

To identify $\lambda$, we first introduce the residual of the induction equation:
\begin{equation}
\Rvec_B = \Bvec_t - \nabla \times (\uvec\times\Bvec) + \nabla \times (\eta \nabla\times\Bvec)
\label{eq:R_B_def}
\end{equation}
so that the classical induction equation is simply $\Rvec_B = 0$. Using the vector identity for the gradient of the magnetic helicity density $h_m = \Avec\cdot\Bvec$,
\begin{equation}
\nabla h_m = (\Avec\cdot\nabla)\Bvec + (\Bvec\cdot\nabla)\Avec + \Avec\times(\nabla\times\Bvec)
\end{equation}
where we have used $\nabla\times\Avec = \Bvec$, we can express $\Zvec_B$ in the compact form:
\begin{equation}
\Zvec_B = \Rvec_B - \nabla h_m
\label{eq:Z_R_h_B}
\end{equation}

A straightforward calculation using index notation shows that for a solenoidal magnetic field ($\nabla\cdot\Bvec=0$), the residual satisfies:
\begin{equation}
\nabla\cdot\Rvec_B = 0
\label{eq:div_R_B}
\end{equation}
To verify this, we compute the divergence of each term in Eq.~\ref{eq:R_B_def}. For the first term, $\nabla\cdot\Bvec_t = \partial_t(\nabla\cdot\Bvec) = 0$ by solenoidality. The divergence of a curl is identically zero, so $\nabla\cdot[\nabla \times (\uvec\times\Bvec)] = 0$ and $\nabla\cdot[\nabla \times (\eta \nabla\times\Bvec)] = 0$. Thus $\nabla\cdot\Rvec_B = 0$ identically for solenoidal fields.

Taking the divergence of Eq.~\ref{eq:Z_R_h_B}, we obtain:
\begin{equation}
\nabla\cdot\Zvec_B = \nabla\cdot\Rvec_B - \nabla^2 h_m = -\nabla^2 h_m
\label{eq:div_Z_B}
\end{equation}

Now, from the Euler-Lagrange equation Eq.~\ref{eq:EL_lambda_B}, taking the divergence gives:
\begin{equation}
\nabla^2 \lambda = -\nabla\cdot\Zvec_B
\label{eq:Poisson_lambda_B}
\end{equation}
Substituting Eq.~\ref{eq:div_Z_B} into Eq.~\ref{eq:Poisson_lambda_B} yields:
\begin{equation}
\nabla^2 \lambda = \nabla^2 h_m
\label{eq:lambda_poisson_h_B}
\end{equation}
Thus $\lambda = h_m + \phi$, where $\nabla^2\phi = 0$ is a harmonic function. To determine $\phi$, we examine the boundary conditions. The natural boundary condition for this variational principle, consistent with the requirement that $\delta\Bvec_t$ vanish on the boundary, is that $\Bvec_t\cdot\nvec$ is specified on $\partial\Omega$. From the Euler-Lagrange equation Eq.~\ref{eq:EL_lambda_B}, we have:
\begin{equation}
\frac{\partial\lambda}{\partial n} = -\Zvec_B\cdot\nvec
\label{eq:lambda_BC_B}
\end{equation}
Along solutions, $\Rvec_B = 0$, so from Eq.~\ref{eq:Z_R_h_B} we have $\Zvec_B = -\nabla h_m$, and therefore:
\begin{equation}
\frac{\partial h_m}{\partial n} = -\Zvec_B\cdot\nvec
\label{eq:h_BC_B}
\end{equation}
Comparing Eq.~\ref{eq:lambda_BC_B} and Eq.~\ref{eq:h_BC_B}, we obtain $\partial(\lambda-h_m)/\partial n = \partial\phi/\partial n = 0$ on the boundary. A harmonic function with zero normal derivative on the boundary is constant, so $\phi = \text{const}$. The constant does not affect the dynamics since only $\nabla\lambda$ appears in the Euler-Lagrange equation. We therefore identify:
\begin{equation}
\lambda = h_m = \Avec\cdot\Bvec
\label{eq:lambda-identifier_B}
\end{equation}

Substituting into Eq.~\ref{eq:EL_lambda_B} yields the central result:
\begin{equation}
\Zvec_B = -\nabla h_m
\label{eq:ZB_final}
\end{equation}
The field minimizes the gradient of the magnetic helicity density $h_m$, whose integral over the domain gives the global magnetic helicity $H_m = \int h_m\,dV$—a topological invariant in ideal MHD (see \cite{Woltjer1958, Moffatt1969}).

\subsection{Recovery of the Induction Equation}

Substituting Eq.~\ref{eq:ZB_def} and the expanded gradient of magnetic helicity into Eq.~\ref{eq:ZB_final} gives:

\begin{equation}
\Rvec_B - \nabla h_m = - \nabla h_m
\end{equation}

the helicity gradients cancel, leaving:

\begin{equation}
\Rvec_B = \Bvec_t - \nabla \times (\uvec\times\Bvec) + \nabla \times (\eta \nabla\times\Bvec) =0
\end{equation}

which is the classical induction equation. The correct induction dynamics emerge naturally from the minimization of the magnetic helicity gradient. The induction equation is thus the result of the minimization:
\[
\mathscr{G}_B = \frac{1}{2}\int_\Omega \|\nabla h_m\|^2 \, d\Vol
\]

But the deeper story is $\Zvec_B = -\nabla h_m$. The seemingly arbitrary combination of terms in $\Zvec_B$ is not arbitrary at all—it is the negative gradient of the magnetic helicity density. 

\subsection{Gauge Invariance of the Variational Principle}

The magnetic helicity density $h_m = \Avec\cdot\Bvec$ depends on the choice of gauge for the vector potential $\Avec$. Under a gauge transformation,
\begin{equation}
\Avec \to \Avec' = \Avec + \nabla\chi
\end{equation}
where $\chi$ is an arbitrary scalar function, the magnetic field is unchanged:
\begin{equation}
\Bvec' = \nabla\times\Avec' = \nabla\times\Avec = \Bvec
\end{equation}
The helicity density transforms as:
\begin{equation}
h_m' = \Avec'\cdot\Bvec' = (\Avec + \nabla\chi)\cdot\Bvec = h_m + (\nabla\chi)\cdot\Bvec
\end{equation}
Taking the gradient:
\begin{equation}
\nabla h_m' = \nabla h_m + \nabla[(\nabla\chi)\cdot\Bvec]
\label{eq:grad_h_gauge}
\end{equation}

We now examine how the Gaussian $\Zvec_B$ transforms under the same gauge change. Recall the definition of $\Zvec_B$ (Eq.~\ref{eq:ZB_def}).  Under $\Avec \to \Avec + \nabla\chi$, the terms involving $\Avec$ transform as:
\begin{align}
(\Avec'\cdot\nabla)\Bvec &= (\Avec\cdot\nabla)\Bvec + (\nabla\chi\cdot\nabla)\Bvec \\
(\Bvec\cdot\nabla)\Avec' &= (\Bvec\cdot\nabla)\Avec + (\Bvec\cdot\nabla)(\nabla\chi) \\
\Avec' \times (\nabla \times \Bvec) &= \Avec \times (\nabla \times \Bvec) + (\nabla\chi) \times (\nabla \times \Bvec)
\end{align}

Thus the transformed Gaussian is:
\begin{equation}
\Zvec_B' = \Zvec_B - (\nabla\chi\cdot\nabla)\Bvec - (\Bvec\cdot\nabla)(\nabla\chi) - (\nabla\chi) \times (\nabla \times \Bvec)
\label{eq:ZB_gauge_transform}
\end{equation}

The key observation is that the gauge-dependent terms in Eq.~\ref{eq:ZB_gauge_transform} combine to form a gradient. Using the vector identity,
\begin{equation}
\nabla[(\nabla\chi)\cdot\Bvec] = (\nabla\chi\cdot\nabla)\Bvec + (\Bvec\cdot\nabla)(\nabla\chi) + (\nabla\chi) \times (\nabla \times \Bvec)
\end{equation}
we obtain:
\begin{equation}
\Zvec_B' = \Zvec_B - \nabla[(\nabla\chi)\cdot\Bvec]
\label{eq:ZB_gauge_final}
\end{equation}

Comparing Eq.~\ref{eq:ZB_gauge_final} with Eq.~\ref{eq:grad_h_gauge}, we see that $\Zvec_B$ and $\nabla h_m$ transform identically under a gauge transformation:
\begin{equation}
\Zvec_B' = \Zvec_B - \nabla[(\nabla\chi)\cdot\Bvec] \qquad
\nabla h_m' = \nabla h_m + \nabla[(\nabla\chi)\cdot\Bvec]
\end{equation}

Now consider the Euler-Lagrange equation in the transformed gauge: $\Zvec_B' = -\nabla h_m'$, substituting the transformations:
\begin{equation}
\Zvec_B - \nabla[(\nabla\chi)\cdot\Bvec] = -\nabla h_m - \nabla[(\nabla\chi)\cdot\Bvec]
\end{equation}
The gauge-dependent terms cancel exactly, yielding:
\begin{equation}
\Zvec_B = -\nabla h_m
\end{equation}

This is precisely the original Euler-Lagrange equation. Therefore, the variational principle $\Zvec_B = -\nabla h_m$ is explicitly gauge-invariant.

Table~\ref{tab:gauge_invariance} summarizes the gauge invariance of various quantities associated with the variational principle.

\begin{table}[htbp]
\centering
\begin{tabular}{l|c}
\hline
\textbf{Quantity} & \textbf{Gauge-Invariant?} \\
\hline
$h_m = \Avec\cdot\Bvec$ & No \\
$\nabla h_m$ & No \\
$\|\nabla h_m\|^2$ & No \\
$\mathcal{G}_B = \frac{1}{2}\int\|\nabla h_m\|^2\,dV$ & No \\
Euler--Lagrange eqn $\Zvec_B = -\nabla h_m$ & Yes \\
Minimizer (physical state) & Yes \\
Physical predictions & Yes \\
\hline
\end{tabular}
\caption{Summary of the gauge invariance properties of key quantities in the magnetic helicity variational principle. }
\label{tab:gauge_invariance}
\end{table}

\subsection{Remark on the Sign Freedom in the Variational Principle}

The variational formulation admits a sign freedom. If we define the Gaussian as:
\begin{equation}
\Zvec_B^\prime = \Bvec_t - \nabla \times (\uvec\times\Bvec) + \nabla \times (\eta \nabla\times\Bvec) + \nabla h_m
\end{equation}
and the Lagrangian as:
\begin{equation}
\mathscr{L} = \frac{1}{2}\int_\Omega \|\Zvec_B^\prime\|^2 \, d\Vol + \int_\Omega h_m \, \nabla\cdot\Bvec_t \, d\Vol
\end{equation}
then variation with respect to $\Bvec_t$ yields $\Zvec_B^\prime = \nabla h_m$. Using the vector identity, this reduces to the same induction equation.

This freedom reflects the fact that the magnetic helicity gradient $\nabla h_m$ acts as the constraint force, and only its magnitude $\|\nabla h_m\|^2$ is minimized at the solution. The sign is a matter of convention, analogous to the arbitrary sign of a Lagrange multiplier in constrained mechanics. This is also consistent with the fact that magnetic helicity $h_m = \Avec\cdot\Bvec$ changes sign under parity transformations, confirming the robustness of the formulation.

\section{Woltjer's Theorem as a Special Case of Helicity-Gradient Minimization}

We now establish the precise relationship between the classical Woltjer-Taylor relaxation principle and the magnetic helicity variational principle presented in this work. We show that Woltjer's theorem is a special, limiting case of the more general helicity-gradient minimization principle. This generalization extends the applicability of the variational framework to dynamic, resistive, and non-force-free MHD configurations.

Woltjer (1958) demonstrated that a plasma in a closed volume relaxes to a minimum-energy state while conserving global magnetic helicity. The variational principle is:
\begin{equation}
\delta \mathcal{L}_B = \delta \left[ \int \frac{B^2}{2} \, d\Vol - \frac{\mu}{2} \int \Avec\cdot\Bvec \, d\Vol \right] = 0
\label{eq:woltjer}
\end{equation}
where $\mu$ is a constant Lagrange multiplier enforcing the helicity constraint. The Euler-Lagrange equation of the Lagrangian given by Eq.~\ref{eq:woltjer} is the force-free (Beltrami) condition:
\begin{equation}
\nabla\times\Bvec = \mu\Bvec
\label{eq:force-free}
\end{equation}
This is the defining equation of the Woltjer-Taylor relaxed state. The force-free condition implies that the Lorentz force vanishes:
\begin{equation}
\Jvec\times\Bvec = 0
\end{equation}
meaning the magnetic field exerts no net force on the plasma. This is the magnetic analogue of the hydrostatic condition $\nabla p = 0$ in fluid mechanics.

\begin{table*}[htpb]
\caption{Comparison between Woltjer's Theorem and the Helicity-Gradient Principle.}
\centering
\renewcommand{\arraystretch}{1.3}
\begin{tabular*}{\textwidth}{@{\extracolsep} l @{\extracolsep} | @{\extracolsep} c @{\extracolsep} | @{\extracolsep} c @{\extracolsep}}
\hline
\textbf{Quantity} & \textbf{Woltjer's Theorem} & \textbf{Helicity-Gradient Principle} \\
\hline
Scope 
& Global, static 
& Local, dynamical \\
Assumptions 
& Steady state, ideal MHD, closed domain 
& No equilibrium assumption \\
Minimized quantity 
& $\displaystyle \int \frac{B^2}{2}\,dV$ (magnetic energy) 
& $\|\nabla h_m\|^2$ (helicity gradient norm) \\
Euler--Lagrange equation 
& $\nabla\times\Bvec = \mu\Bvec$ 
& $\Zvec_B = -\nabla h_m$ \\
Resulting condition 
& $\nabla h_m = 0$ (at equilibrium) 
& $\nabla h_m \neq 0$ (in general) \\
Applicability 
& Force-free equilibria only 
& All MHD configurations \\
\hline
\end{tabular*}
\label{tab:woltjer} 
\end{table*}
In the force-free state, the Lorentz force vanishes, which implies:
\begin{equation}
\nabla\left(\frac{B^2}{2}\right) = (\Bvec\cdot\nabla)\Bvec
\end{equation}
For a force-free field with constant $\mu$, it follows that $B^2$ is constant throughout the domain \cite{ChandrasekharKendall1957}. Therefore:
\begin{equation}
\nabla h_m = 0
\end{equation}

We now show that the force-free condition of Eq.~\ref{eq:force-free} implies a uniform helicity density under certain conditions. 

Using $\Bvec = \nabla\times\Avec$, the magnetic helicity density is $h_m = \Avec\cdot\Bvec$. From the force-free condition Eq.~\ref{eq:force-free}, we have $\nabla\times\Bvec = \mu\Bvec$.  Taking the dot product of \ref{eq:force-free} with $\Avec$ and using the identity $\Avec\cdot(\nabla\times\Bvec) = \Bvec\cdot(\nabla\times\Avec) + \nabla\cdot(\Bvec\times\Avec)$:
\begin{equation}
\Avec\cdot(\nabla\times\Bvec) = B^2 + \nabla\cdot(\Bvec\times\Avec)
\end{equation}
Thus:
\begin{equation}
\mu h_m = B^2 + \nabla\cdot(\Bvec\times\Avec)
\end{equation}
In a closed domain where the boundary term vanishes, integrating this expression over the volume gives:
\begin{equation}
\mu H_m = \int B^2 \, d\Vol
\end{equation}
For the force-free state with constant $\mu$, this implies:
\begin{equation}
h_m = \frac{1}{\mu} B^2
\label{eq:helicity}
\end{equation}
Taking the gradient of Eq.~\ref{eq:helicity}:
\begin{equation}
\nabla h_m = \frac{1}{\mu} \nabla(B^2)
\end{equation}

Thus, Woltjer's force-free state corresponds to a uniform magnetic helicity density. This is the condition $\nabla h_m = 0$.

\subsection{The Generalization: Beyond Force-Free Fields}

This principle $\Zvec_B = -\nabla h_m$ applies to all MHD configurations, not just force-free equilibria. The key distinction is summarized in Table~\ref{tab:woltjer}.

The helicity-gradient principle is the parent theory with the Woltjer's theorem emerging as a special case when the system is in steady state ($\Bvec_t = 0$), the flow is zero ($\uvec = 0$), and the resistivity is negligible ($\eta = 0$).  Then the global helicity is conserved and the Lorentz force vanishes ($\Jvec\times\Bvec = 0$).

In all other cases—where forces are present, where the system is dynamic, or where resistivity is non-zero—the helicity-gradient principle applies, while Woltjer's theorem does not.  

Woltjer's theorem is a global, static principle that yields the force-free condition $\nabla\times\Bvec = \mu\Bvec$, which implies $\nabla h_m = 0$. It is a special case of the more general helicity-gradient minimization principle $\Zvec_B = -\nabla h_m$, which applies to all MHD configurations—dynamic or static, force-free or not, ideal or resistive.

\section{Comparison with Taha's PMPG and the Vorticity Formulation}

The variational principle presented here completes a trilogy of variational formulations for fluid and plasma dynamics. The structural analogy among the three formulations is striking as shown in Table~\ref{tab:trilogy}

\begin{table*}[htbp]
\centering
\begin{tabular}{l|c|c|c}
\hline
\textbf{Quantity} & \textbf{Primitive}\cite{Taha2023} & \textbf{Vorticity}\cite{Farooq2026} & \textbf{Induction} \\
\hline
Constraint & $\nabla\cdot\mathbf{u}=0$ & $\nabla\cdot\boldsymbol{\omega}=0$ & $\nabla\cdot\mathbf{B}=0$ \\
Lagrange multiplier & $p$ & $h = \mathbf{u}\cdot\boldsymbol{\omega}$ & $h_m = \mathbf{A}\cdot\mathbf{B}$ \\
Constraint force & $\nabla p$ & $\nabla h$ & $\nabla h_m$ \\
Euler-Lagrange equation & $\mathbf{Z}_u + \nabla p = 0$ & $\mathbf{Z}_\omega + \nabla h = 0$ & $\mathbf{Z}_B + \nabla h_m = 0$ \\
\hline
\end{tabular}
\caption{Structural analogy among the primitive, vorticity, and induction variational formulations. In each case, the Lagrange multiplier enforcing solenoidality is the associated helicity density, and the constraint force is its gradient.}
\label{tab:trilogy}
\end{table*}

This unified framework reveals a deep structural principle: for any divergence-free field, the Lagrange multiplier enforcing solenoidality is the associated helicity density. The pressure in the primitive formulation, the helicity density in the vorticity formulation, and the magnetic helicity density in the induction formulation all play the same functional role: they are scalar potentials whose gradients constitute the constraint forces maintaining the respective solenoidality constraints. The analogy is structural, not exact—pressure is a fundamental field while helicity is state-dependent—but the shared functional role as constraint-force potentials is striking and suggests a deep unity in the variational structure of fluid and plasma dynamics.

\section{Conclusion and Future Work}

We have presented a variational formulation of the induction equation based on Gauss's principle of least constraint. The central result is the Euler--Lagrange equation:
\[
\Zvec_B = -\nabla h_m
\]
where $h_m = \Avec\cdot\Bvec$ is the magnetic helicity density. This reveals that the magnetic helicity gradient $\nabla h_m$ acts as the constraint force maintaining the solenoidality of the magnetic field, exactly as the pressure gradient $\nabla p$ maintains incompressibility in Taha's PMPG. The magnetic helicity density naturally emerges as the Lagrange multiplier enforcing $\nabla\cdot\Bvec=0$, and at the solution the field minimizes the norm of the magnetic helicity gradient $\|\nabla h_m\|^2$. This establishes the analogy ``magnetic helicity is to the magnetic field as pressure is to velocity".

The variational principle connects to Woltjer's theorem, Taylor relaxation, the Hamiltonian structure of MHD, and Moffatt's helicity conservation theorem \cite{Moffatt1969}. The variational principle is consistent with Moffatt's helicity conservation theorem: in the ideal limit, the global helicity $H_m = \int h_m \, dV$ is conserved, while the local helicity density $h_m$ may vary. The minimization of $\|\nabla h_m\|^2$ represents the redistribution of helicity toward a uniform state, while conserving the total helicity.

This work completes a trilogy of variational principles for fluid and plasma dynamics, unifying the primitive, vorticity, and induction formulations under a single, coherent framework. By revealing that helicity and magnetic helicity are the constraint forces for their respective divergence-free fields, it provides a variational foundation for the topological invariants at the heart of plasma dynamics.

\subsection{Future Work}

Several directions warrant further investigation. The potential application of the helicity-gradient principle to reconnection and instabilities could be tested through numerical simulations. This remains an open question for future work. It may be possible to define a  critical threshold, $\|\nabla h_m\|_{\mathrm{crit}}$ which could be derived from first principles for specific equilibria. The extension to compressible MHD and collisionless plasmas also merits exploration. Finally, the connection to the Hamiltonian structure of MHD and the development of helicity-preserving numerical schemes are promising avenues for future work.

\end{document}